\documentclass[fleqn,12pt]{article}

\usepackage{latexsym,ifthen,epsfig,color}
\usepackage{amsmath,amssymb,amsthm}
\usepackage{float}

\newcommand{\join}{\text{\textcircled{{\footnotesize 1}}}}
\newcommand{\cojoin}{\text{\textcircled{{\footnotesize 0}}}}

\newcommand{\NP}{\ensuremath{\mathbb{NP}}}

\newtheorem{clai}{Claim}[section]
\newtheorem{theo}{Theorem}
\newtheorem{lemma}{Lemma}

\newtheorem{coro}{Corollary}

\begin{document}

\author{
Andreas Brandst\"adt\\
\small Institut f\"ur Informatik, Universit\"at Rostock, D-18051 Rostock, Germany\\
\small \texttt{andreas.brandstaedt@uni-rostock.de}\\
\and
Raffaele Mosca\\
\small Dipartimento di Economia, Universit\'a degli Studi ``G. D'Annunzio'', 
Pescara 65121, Italy\\
\small \texttt{r.mosca@unich.it}
}

\title{Finding Efficient Domination for $S_{1,3,3}$-Free Bipartite Graphs in Polynomial Time}

\maketitle

\begin{abstract}
A vertex set $D$ in a finite undirected graph $G$ is an {\em efficient dominating set} (\emph{e.d.s.}\ for short) of $G$ if every vertex of $G$ is dominated by exactly one vertex of $D$. The \emph{Efficient Domination} (ED) problem, which asks for the existence of an e.d.s.\ in $G$, is \NP-complete for various $H$-free bipartite graphs, e.g., Lu and Tang showed that ED is \NP-complete for chordal bipartite graphs and for planar bipartite graphs; actually, ED is \NP-complete even for planar bipartite graphs with vertex degree at most 3 and girth at least $g$ for every fixed $g$. Thus, ED is \NP-complete for $K_{1,4}$-free bipartite graphs and for $C_4$-free bipartite graphs. 

In this paper, we show that ED can be solved in polynomial time for $S_{1,3,3}$-free bipartite graphs. 
\end{abstract}

\noindent{\small\textbf{Keywords}:
Efficient domination;
$S_{1,3,3}$-free bipartite graphs.
}

\section{Introduction}\label{sec:intro}

Let $G=(V,E)$ be a finite undirected graph. A vertex $v$ {\em dominates} itself and its neighbors. A vertex subset $D \subseteq V$ is an {\em efficient dominating set} ({\em e.d.s.}\ for short) of $G$ if every vertex of $G$ is dominated by exactly one vertex in $D$; for any e.d.s.\ $D$ of $G$, $|D \cap N[v]| = 1$ for every $v \in V$ (where $N[v]$ denotes the closed neighborhood of $v$).
Note that not every graph has an e.d.s.; the {\sc Efficient Dominating Set} (ED for short) problem asks for the existence of an e.d.s.\ in a given graph~$G$.

\medskip

In \cite{BanBarSla1988,BanBarHosSla1996}, it was shown that the ED problem is \NP-complete, and it is \NP-complete for $P_k$-free graphs, $k \ge 7$, i.e., 
ED is \NP-complete for $S_{1,3,3}$-free graphs. 
However, in \cite{BraMos2016}, we have shown that ED is solvable in polynomial time for $P_6$-free graphs which leads to a dichotomy of ED for $H$-free graphs. 

\medskip

Lu and Tang \cite{LuTan2002} showed that ED is \NP-complete for chordal bipartite graphs (i.e., hole-free bipartite graphs). Thus, for every $k \ge 3$, ED is \NP-complete for $C_{2k}$-free bipartite graphs. 
Moreover, ED is \NP-complete for planar bipartite graphs \cite{LuTan2002} and even for planar bipartite graphs of maximum degree 3 \cite{BraMilNev2013} and girth at least $g$ for every fixed $g$ \cite{Nevri2014}. Thus, ED is \NP-complete for $K_{1,4}$-free bipartite graphs and for $C_4$-free bipartite graphs.

\medskip

It is well known that for every graph class with bounded clique-width, ED can be solved in polynomial time \cite{CouMakRot2000}; for instance, the clique-width of  claw-free bipartite, i.e., $K_{1,3}$-free bipartite graphs is bounded. Dabrowski and Paulusma  \cite{DabPau2016} published a dichotomy for clique-width of $H$-free bipartite graphs. For instance, the clique-width of $S_{1,2,3}$-free bipartite graphs is bounded (which includes $K_{1,3}$-free bipartite graphs). 
However, the clique-width of $S_{1,3,3}$-free bipartite graphs is unbounded. 

\medskip

In \cite{BraMos2019}, we solved ED in polynomial time for $P_7$-free bipartite graphs, for $\ell P_4$-free bipartite graphs for fixed $\ell$, for $S_{2,2,4}$-free bipartite graphs as well as for $P_9$-free bipartite graphs with degree at most 3, but we had some open problems: 
What is the complexity of ED for 
\begin{itemize}
\item[$-$] $P_k$-free bipartite graphs, $k \ge 8$, 
\item[$-$] $S_{1,3,3}$-free bipartite graphs, 
\item[$-$] $S_{1,1,5}$-free bipartite graphs,
\item[$-$] $S_{2,2,k}$-free bipartite graphs for $k \ge 5$,
\item[$-$] chordal bipartite graphs with vertex degree at most 3? 
\end{itemize}

In \cite{BraMos2021/2}, we have shown already that for $S_{1,1,5}$-free bipartite graphs, ED is solvable in polynomial time, and in 
\cite{BraMos2021/3}, we have shown already that for $P_8$-free bipartite graphs, ED is solvable in polynomial time.
In this manuscript, we will show:

\begin{theo}\label{EDS133frbipgr}
For $S_{1,3,3}$-free bipartite graphs, ED is solvable in polynomial time.
\end{theo}

\section{Preliminaries}

Let $G=(X,Y,E)$ with $V(G)=X \cup Y$, be an $S_{1,3,3}$-free bipartite graph, say, every vertex in $X$ is black, and every vertex in $Y$ is white. 
For subsets $U,W \in V(G)$, the {\em join} $U \join W$ denotes $uw \in E$ for every $u \in U$ and every $w \in W$. Moreover, the {\em cojoin} $U \cojoin W$ denotes
$uw \notin E$ for every $u \in U$ and every $w \in W$. In particular, for $u \in V(G)$, $u \join W$ if $uw \in E$ for every $w \in W$ and $u \cojoin W$ if $uw \notin E$ for every $w \in W$.

\medskip

Let $P_i$, $i \ge 1$, denote the chordless path with $i$ vertices, and in bipartite graphs, let $C_{2i}$, $i \ge 2$, denote the chordless cycle with $2i$ vertices. 
Recall that a subgraph $2P_i$, $i \ge 3$, has two $P_i$'s $P$, $P'$ without any contact between $P$ and $P'$, i.e., $P \cojoin P'$. For example, if $(u_1,v_1,w_1)$, $(u_2,v_2,w_2)$ induce a $2P_3$ in $G$ then $\{u_1,v_1,w_1\} \cojoin \{u_2,v_2,w_2\}$. 

\medskip

For indices $i,j,k \ge 0$, let $S_{i,j,k}$ denote the graph with vertices $u,r_1,\ldots,r_i$, $s_1,\ldots,s_j$, $t_1,\ldots,t_k$ (and {\em midpoint} $u$) such that the subgraph induced by $u,r_1,\ldots,r_i$ forms a $P_{i+1}$ ($u,r_1,\ldots,r_i$), the subgraph induced by $u,s_1,\ldots,s_j$ forms a $P_{j+1}$ ($u,s_1,\ldots,s_j$), and the subgraph induced by $u,t_1,\ldots,t_k$ forms a $P_{k+1}$ ($u,t_1,\ldots,t_k$), and there are no other edges in $S_{i,j,k}$. Thus, claw is $S_{1,1,1}$, chair is $S_{1,1,2}$, and $P_k$ is isomorphic to $S_{0,0,k-1}$.

\medskip

Clearly, an $S_{1,3,3}$ $S$ in $G$ has eight vertices, say $(u_1,v_1,u_2,v_2,u_3,v_3,u_4)$ induce a $P_7$ with midpoint $v_2$ and $v_2u \in E$ with $uv_1 \notin E$, 
$uv_3 \notin E$, i.e., $(v_2,u,u_2,v_1,u_1,u_3,v_3,u_4)$ (with midpoint $v_2$) induce an $S_{1,3,3}$ in $G$.  
The bipartite graph $G$ is {\em $S_{1,3,3}$-free} if there is no such $S_{1,3,3}$ in $G$.  

\medskip

Clearly, in general, $S_{1,3,3}$-free bipartite graphs are not $P_k$-free for any $k \ge 1$; for example, every bipartite graph $G=P_k$, $k \ge 1$, is $S_{1,3,3}$-free.  

\medskip

Let $D$ be a possible e.d.s.\ of $G$. 
For example, if $G=P_{1}$, say $(r_1)$, then $D=\{r_1\}$, if $G=P_{2}$, say $(r_1,r_2)$, then $D=\{r_1\}$ or $D=\{r_2\}$, if $G=P_{3}$, say $(r_1,r_2,r_3)$,  then $D=\{r_2\}$, if $G=P_{4}$, say $(r_1,r_2,r_3,r_4)$, then $D=\{r_1,r_4\}$ etc. For instance, if  
$G=P_{10}$, say $(r_1,\ldots,r_{10})$, then $D=\{r_1,r_4,r_7,r_{10}\}$. 
In general, for every $G=P_k$, $k \ge 1$, $G$ is $S_{1,3,3}$-free bipartite and has an e.d.s.\ which can be solved in polynomial time.   

Another example is $G=C_{2k}$, $k \ge 2$; every such $G=C_{2k}$ is $S_{1,3,3}$-free bipartite. Clearly, $G=C_4$ has no e.d.s., $G=C_6$ has three e.d.s., 
$G=C_8$ has no e.d.s., $G=C_{10}$ has no e.d.s., $G=C_{12}$ has three e.d.s.\ etc. In general, $G=C_{2k}$, $k \ge 2$, has an e.d.s.\ if and only if $k=3i$, $i \ge 1$. 
   
\medskip

A vertex $u \in V(G)$ is {\em forced} if $u \in D$ for every e.d.s.\ $D$ of $G$; $u$ is {\em excluded} if $u \notin D$ for every e.d.s.\ $D$ of $G$.
For example, if $x_1,x_2 \in X$ are leaves in $G$ and $y$ is the neighbor of $x_1,x_2$ then $x_1,x_2$ are excluded and $y$ is forced.

\medskip

By a forced vertex, $G$ can be reduced to $G'$ as follows:

\begin{clai}\label{forcedreduction}
If $u$ is forced then $G$ has an e.d.s.\ $D$ with $u \in D$ if and only if the reduced graph $G'=G \setminus N[u]$ has an e.d.s.\ $D'=D \setminus \{u\}$ such that all vertices in $N^2(u)$ are excluded in $G'$.
\end{clai}

Analogously, if we assume that $v \in D \cap V(G)$ then $u \in V(G)$ is {\em $v$-forced} if $u \in D$ for every e.d.s.\ $D$ of $G$ with $v \in D$,
and $u$ is {\em $v$-excluded} if $u \notin D$ for every e.d.s.\ $D$ of $G$ with $v \in D$. For checking whether $G$ has an e.d.s.\ $D$ with $v \in D$, we can clearly reduce $G$ by forced vertices as well as by $v$-forced vertices when we assume that $v \in D$:

\begin{clai}\label{vforcedreduction}
If $v \in D$ and $u$ is $v$-forced then $G$ has an e.d.s.\ $D$ with $v \in D$ if and only if the reduced graph $G'=G \setminus N[u]$ has an e.d.s.\ $D'=D \setminus \{u\}$ with $v \in D'$ such that all vertices in $N^2(u)$ are $v$-excluded in $G'$.
\end{clai}

Similarly, for $k \ge 2$, $u \in V(G)$ is {\em $(v_1,\ldots,v_k)$-forced} if $u \in D$ for every e.d.s.\ $D$ of $G$ with $v_1,\ldots,v_k \in D$, and correspondingly, $u \in V(G)$ is {\em $(v_1,\ldots,v_k)$-excluded} if $u \notin D$ for such e.d.s.\ $D$, and $G$ can be reduced by the same principle.

\medskip

Clearly, for every connected component of $G$, the e.d.s.\ problem is independently solvable. Thus, we can assume that $G$ is connected.   
For every vertex $v \in V(G)$, either $v \in D$ or $|N(v) \cap D|=1$. 

\medskip

If for an e.d.s.\ $D$ in the bipartite graph $G=(X,Y,E)$, $|D|=1$, say without loss of generality, $D =\{x\}$ with $x \in X$ and 
$D \cap Y=\emptyset$, then every $y \in Y$ must have the $D$-neighbor $x \in D$, i.e., $x \join Y$ and by the e.d.s.\ property, $|X|=1$, i.e., $X=\{x\}$ (else there is no such e.d.s.\ in $G$), which is a trivial e.d.s.\ solution. 

\medskip

Now assume that $|D| \ge 2$, and without loss of generality, $D \cap X \neq \emptyset$. 
Since $G$ is connected, every $x \in D \cap X$ must have at least distance $2$ in $G$, say $(x,y,x')$ induce a $P_3$ in $G$. Then by the e.d.s.\ property, 
$y,x' \notin D$ and $x'$ must have a $D$-neighbor $y' \in D \cap Y$ (else there is no such e.d.s.\ in $G$). Thus, $D \cap X \neq \emptyset$ and 
$D \cap Y \neq \emptyset$, say $(x,y,x',y')$ induce a $P_4$ in $G$ with $x,y' \in D$.   

Recall that ED is solvable in polynomial time for $P_7$-free bipartite graphs \cite{BraMos2019}. Now there are $P_7$'s in $G$. 
If $(u_1,v_1,u_2,v_2,u_3,v_3,u_4)$ induce a $P_7$ in $G$ and $v_1,v_3 \in D$ then by the e.d.s.\ property, $u_2,v_2,u_3 \notin D$ and $v_2$ must have a $D$-neighbor $u \in D$. But then $(v_2,u,u_2,v_1,u_1,u_3,v_3,u_4)$ (with midpoint $v_2$) induce an $S_{1,3,3}$ in $G$, which is a contradiction. Thus, either $v_1 \notin D$ or   
$v_3 \notin D$. Moreover, if $dist_G(v,v')=4$ with $v,v' \in D$ then there is no such $P_7$ with $v,v' \in D$. 

\subsection{When no $D$-vertex is midpoint of a $P_5$ in $G$}\label{noDvertexmidpointP5}

In this subsection, every $D$-vertex is no midpoint of a $P_5$ in $G$.

\begin{clai}\label{P7Pu2v2u3excluded}
For every $P_7$ $P=(u_1,v_1,u_2,v_2,u_3,v_3,u_4)$ in $G$, $u_2,v_2,u_3$ are excluded. 
\end{clai}

\noindent
{\em Proof.}
Let $P=(u_1,v_1,u_2,v_2,u_3,v_3,u_4)$ induce a $P_7$ in $G$ and recall that in this subsection,   
every $D$-vertex is no midpoint of a $P_5$ in $G$.

\medskip

Since $(u_1,v_1,u_2,v_2,u_3)$ induce a $P_5$ in $G$ and $u_2$ is midpoint of the $P_5$, $u_2 \notin D$. Analogously, 
since $(v_1,u_2,v_2,u_3,v_3)$ induce a $P_5$ in $G$ and $v_2$ is midpoint of the $P_5$, $v_2 \notin D$, and 
since $(u_2,v_2,u_3,v_3,u_4)$ induce a $P_5$ in $G$ and $u_3$ is midpoint of the $P_5$, $u_3 \notin D$, i.e., $u_2,v_2,u_3$ are excluded 
 and Claim \ref{P7Pu2v2u3excluded} is shown.
\qed

\begin{coro}\label{DnomidpointP5noP7}
In this subsection, there are no such $P_7$'s $P=(u_1,v_1,u_2,v_2,u_3,v_3)$ in $G$. 
\end{coro}

\noindent
{\em Proof.}
Let $P=(u_1,v_1,u_2,v_2,u_3,v_3,u_4)$ induce a $P_7$ in $G$ and recall that by Claim \ref{P7Pu2v2u3excluded}, $u_2,v_2,u_3$ are excluded. 
Then $v_2$ must have a $D$-neighbor $u \in D$, $u \neq u_2,u_3$. Since in this subsection, $u \in D$ is no midpoint of a $P_5$ in $G$, 
$uv_1 \notin E$ (else $(u_1,v_1,u,v_2,u_3)$ induce a $P_5$ with midpoint $u \in D$) and 
$uv_3 \notin E$ (else $(u_2,v_2,u,v_3,u_4)$ induce a $P_5$ with midpoint $u \in D$). 
But then $(v_2,u,u_2,v_1,u_1,u_3,v_3,u_4)$ (with midpoint $v_2$) induce an $S_{1,3,3}$ in $G$, which is a contradiction.  

Thus, there are no such $P_7$'s $P=(u_1,v_1,u_2,v_2,u_3,v_3)$ in $G$, and Corollary \ref{DnomidpointP5noP7} is shown.
\qed

\medskip

Since by Corollary \ref{DnomidpointP5noP7}, there are no such $P_7$'s in $G$ then ED is solvable in polynomial time for $P_7$-free bipartite graphs \cite{BraMos2019}. 
Now assume that there are $D$-vertices which are midpoints of a $P_5$ in $G$. 

\section{Distance levels $N_i$, $i \ge 0$}

Let $N_0:=D_{basis}$, and for every $D_{basis}$-forced vertex $d \in D$, $d \in D_{basis}$, i.e., $D_{basis}:= D_{basis} \cup \{d\}$. 
Let $N_i$, $i \ge 1$, be the distance levels of $D_{basis}$. 

Recall that $D \cap X \neq \emptyset$ and $D \cap Y \neq \emptyset$, say $(x,y_1,x_1,y)$ induce a $P_4$ with $x,y \in D$. Assume that $N_0 \cap X \neq \emptyset$ and $N_0 \cap Y \neq \emptyset$, say $x \in N_0 \cap X$ and $y \in N_0 \cap Y$ with $dist_G(x,y) = 3$, say $(x,y_1,x_1,y)$ induce a $P_4$ in $G$ with $y_1,x_1 \in N_1$, and every $(x,y)$-forced vertex is in $D_{basis}$. 
By the e.d.s.\ property, we have: 
\begin{equation}\label{DN1N2empty}
D \cap (N_1 \cup N_2)=\emptyset 
\end{equation}

If there is a $u \in N_2$ with $N(u) \cap N_3=\emptyset$ then by the e.d.s.\ property, there is no such e.d.s.\ $D$ in $G$ with $N_0=D_{basis}$. Thus, we assume:
\begin{equation}\label{uinN2neighbinN3}
\mbox{ Every vertex } u \in N_2 \mbox{ has a $D$-neighbor in } N_3.
\end{equation}
If $|N(u) \cap N_3| = 1$, say $N(u) \cap N_3=\{v\}$ then $v$ is $D_{basis}$-forced, and one can update $D_{basis}:= D_{basis} \cup \{v\}$ and redefine the distance levels $N_i$, $i \ge 1$, with respect to $D_{basis}$ as above. Now assume:
\begin{equation}\label{uinN2twoneighbinN3}
\mbox{ For every } u \in N_2, |N(u) \cap N_3| \ge 2.
\end{equation}
If for $v \in N_3$, $N(v) \cap (N_3 \cup N_4)=\emptyset$ then by (\ref{DN1N2empty}) and the e.d.s.\ property, $v$ is $D_{basis}$-forced. 
Moreover, if there are two such $v_i \in N_3$, $N(v_i) \cap (N_3 \cup N_4)=\emptyset$, $i \in \{1,2\}$, with common neighbor $u \in N_2$, i.e., 
$uv_i \in E$, $i \in \{1,2\}$, then by the e.d.s.\ property, there is no such e.d.s.\ $D$ in $G$ with $N_0=D_{basis}$. 
Thus assume that such $D_{basis}$-forced vertices $v_i \in N_3$, $i \in \{1,2\}$, do not have any common neighbor in $N_2$. 

Then for every $v \in N_3$ with $N(v) \cap (N_3 \cup N_4)=\emptyset$, one can update $D_{basis}:= D_{basis} \cup \{v\}$ and redefine the distance levels 
$N_i$, $i \ge 1$, with respect to $D_{basis}$ as above. 
Thus, we assume:
\begin{equation}\label{vinN3neighbinN3N4}
\mbox{ Every } v \in N_3 \mbox{ has a neighbor in } N_3 \cup N_4.
\end{equation}

In particular, if $N_4=\emptyset$ then every $v \in N_3$ is $D_{basis}$-forced, and it leads to a simple e.d.s.\ solution in polynomial time or a contradiction if there is an edge in $N_3$ or two vertices in $N_3$ have a common neighbor in $N_2 \cup N_3$. Thus, we can assume:    
\begin{equation}\label{N4nonempty}
N_4 \neq \emptyset.
\end{equation}

\begin{lemma}\label{N5emptyN4indep}
$N_5=\emptyset$ and $N_4$ is independent.
\end{lemma}

\noindent
{\bf Proof.}
Let $(x,y_1,x_1,y)$ induce a $P_4$ in $G[N_0 \cup N_1]$ with $x,y \in N_0$ and $x_1,y_1 \in N_1$. 

\medskip

Suppose to the contrary that $N_5 \neq \emptyset$, say $(r_2,r_3,r_4,r_5)$ induce a $P_4$ with $r_i \in N_i$, $2 \le i \le 5$. Without loss of generality, 
$r_2 \in N_2 \cap X$. Recall that by (\ref{uinN2twoneighbinN3}), $|N(r_2)| \ge 2$, say $r_3,r'_3 \in N_3 \cap N(r_2)$. 

\medskip

First assume that $r_2y_1 \in E$. If $r'_3r_4 \notin E$ then $(r_2,r'_3,r_3,r_4,r_5,y_1,x_1,y)$ (with midpoint $r_2$) induce an $S_{1,3,3}$ in $G$, 
which is a contradiction. Thus, $r'_3r_4 \in E$, i.e., $N(r_2) \cap N_3 \subset N(r_4) \cap N_3$.
Without loss of generality, $r_3 \in D \cap N_3$. Then by the e.d.s.\ property, $r'_3,r_5 \notin D$ and $r_5$ must have a $D$-neighbor in 
$D \cap (N_4 \cup N_5 \cup N_6)$ as well as $r'_3$ must have a $D$-neighbor in $D \cap (N_3 \cup N_4)$.   

If $r'_3$ and $r_5$ have a common $D$-neighbor $d \in D \cap N_4$ then $(r_2,r_3,r'_3,d,r_5,y_1,x_1,y)$ (with midpoint $r_2$) induce an $S_{1,3,3}$ in $G$, which is a contradiction. Thus, $r'_3$ and $r_5$ do not have any common $D$-neighbor, i.e., $r'_3d \in E$ and $r_5d' \in E$, $d,d' \in D$, $d \neq d'$.  
But then $(r'_3,d,r_4,r_5,d',r_2,y_1,x)$ (with midpoint $r'_3$) induce an $S_{1,3,3}$ in $G$, which is a contradiction. 
Thus, $r_2y_1 \notin E$.

\medskip

Next assume that $r_2y'_1 \in E$ with $xy'_1 \in E$, $y_1 \neq y'_1$. Then again, if $r'_3r_4 \notin E$ then $(r_2,r'_3,r_3,r_4,r_5,y'_1,x,y_1)$ (with midpoint $r_2$) induce an $S_{1,3,3}$ 
in $G$, which is a contradiction. Thus, $r'_3r_4 \in E$, i.e., $N(r_2) \cap N_3 \subset N(r_4) \cap N_3$.
Without loss of generality, $r_3 \in D \cap N_3$. Then by the e.d.s.\ property, $r'_3,r_5 \notin D$ and $r_5$ must have a $D$-neighbor in 
$D \cap (N_4 \cup N_5 \cup N_6)$ as well as $r'_3$ must have a $D$-neighbor in $D \cap (N_3 \cup N_4)$.   

If $r'_3$ and $r_5$ have a common $D$-neighbor $d \in D \cap N_4$ then $(r_2,r_3,r'_3,d,r_5,y'_1,x,y_1)$ (with midpoint $r_2$) induce an $S_{1,3,3}$ in $G$, which is a contradiction. Thus, $r'_3$ and $r_5$ do not have any common $D$-neighbor, i.e., $r'_3d \in E$ and $r_5d' \in E$, $d,d' \in D$, $d \neq d'$.  
But then $(r'_3,d,r_4,r_5,d',r_2,y'_1,x)$ (with midpoint $r'_3$) induce an $S_{1,3,3}$ in $G$, which is a contradiction. 
Thus, $r_2y'_1 \notin E$ and $r_2$ does not contact $N(x)$, i.e., $r_2 \cojoin N(x)$. 

\medskip

Next assume that $r_1r_2 \in E$ with $r_1 \in N_1$, $r_1x \notin E$ and $r_0r_1 \in E$ with $r_0 \in N_0$, $r_0 \neq x$.
Since $(r_1,r_0,x_1,y_1,x,r_2,r_3,r_4)$ (with midpoint $r_1$) does not induce an $S_{1,3,3}$ in $G$, $r_1x_1 \notin E$.
If $r_1x'_1 \in E$ with $x'_1y \in E$ then $(r_1,r_0,x'_1,y,x_1,r_2,r_3,r_4)$ (with midpoint $r_1$) induce an $S_{1,3,3}$ in $G$, which is a contradiction. 
Thus, $r_1$ does not contact $N(y)$, i.e., $r_1 \cojoin N(y)$. 

\medskip

Assume without loss of generality that $dist_G(r_0,y)=3$, say $(r_0,r'_1,x'_1,y)$ induce a $P_4$ in $G[N_0 \cup N_1]$ (possibly $x'_1=x_1$).

First assume that $r'_1r_2 \in E$. If $x'_1 \neq x_1$ then, since $(r'_1,r_0,x'_1,y,x_1,r_2,r_3,r_4)$ (with midpoint $r'_1$) does not induce an $S_{1,3,3}$ in $G$, 
$r'_1x_1 \in E$. 

But then $(r'_1,r_0,x_1,y_1,x,r_2,r_3,r_4)$ (with midpoint $r'_1$) induce an $S_{1,3,3}$ in $G$, which is a contradiction. 
Thus, $r'_1r_2 \notin E$. 

If $r'_3r_4 \notin E$ then $(r_2,r'_3,r_3,r_4,r_5,r_1,r_0,r'_1)$ (with midpoint $r_2$) induce an $S_{1,3,3}$ in $G$, 
which is a contradiction. Thus, $r'_3r_4 \in E$, i.e., $N(r_2) \cap N_3 \subset N(r_4) \cap N_3$.
Without loss of generality, $r_3 \in D \cap N_3$. Then by the e.d.s.\ property, $r'_3,r_5 \notin D$ and $r_5$ must have a $D$-neighbor in 
$D \cap (N_4 \cup N_5 \cup N_6)$ as well as $r'_3$ must have a $D$-neighbor in $D \cap (N_3 \cup N_4)$.   

If $r'_3$ and $r_5$ have a common $D$-neighbor $d \in D \cap N_4$ then $(r_2,r_3,r'_3,d,r_5,r_1,r_0,r'_1)$ (with midpoint $r_2$) induce an $S_{1,3,3}$ in $G$, 
which is a contradiction. Thus, $r'_3$ and $r_5$ do not have any common $D$-neighbor, i.e., $r'_3d \in E$ and $r_5d' \in E$, $d,d' \in D$, $d \neq d'$.  
But then $(r'_3,d,r_4,r_5,d',r_2,r_1,r_0)$ (with midpoint $r'_3$) induce an $S_{1,3,3}$ in $G$, which is a contradiction. 
Thus, $N_5=\emptyset$. Analogously, $N_4$ is independent, and Lemma \ref{N5emptyN4indep} is shown.
\qed

\begin{lemma}\label{atmostoneDN3XorY}
$|D \cap N_3 \cap X| \le 1$ and $|D \cap N_3 \cap Y| \le 1$. 
\end{lemma}

\noindent
{\bf Proof.}
Recall that $x \in N_0 \cap X$ and $y \in N_0 \cap Y$ with $dist_G(x,y) = 3$, say $(x,y_1,x_1,y)$ induce a $P_4$ in $G$ with $y_1,x_1 \in N_1$, and every $(x,y)$-forced vertex is in $N_0=D_{basis}$. Then by (\ref{vinN3neighbinN3N4}), for every $D$-vertex $v \in N_3$ which is no such neighbor in $N_3 \cup N_4$, $v$ is  
$D_{basis}$-forced, i.e., $D_{basis}:= D_{basis} \cup \{v\}$.  

\medskip

Suppose to the contrary that there are two $D$-vertices in $N_3 \cap X$ or in $N_3 \cap Y$; without loss of generality, $r_3,s_3 \in D \cap N_3 \cap X$, 
$r_3 \neq s_3$. Recall that by (\ref{vinN3neighbinN3N4}), $r_3$ must have a neighbor $r_4 \in N_3 \cup N_4$, and $s_3$ must have a neighbor $s_4 \in N_3 \cup N_4$.   

Then $(r_2,r_3,r_4)$ induce a $P_3$ with $r_2 \in N_2$ and $r_4 \in N_3 \cup N_4$ as well as $(s_2,s_3,s_4)$ induce a $P_3$ with $s_2 \in N_2$ and 
$s_4 \in N_3 \cup N_4$, i.e., by the e.d.s.\ property, there is a $2P_3$ $(r_2,r_3,r_4)$, $(s_2,s_3,s_4)$ in $G$ with $r_3,s_3 \in D \cap N_3 \cap X$, 
$r_2,s_2 \in N_2 \cap Y$, $r_4,s_4 \in (N_3 \cup N_4) \cap Y$. 

If $r_2$ and $s_2$ have a common neighbor $r_1 \in N_1$ then $r_0r_1 \in E$ with $r_0 \in N_0$. 
But then $(r_1,r_0,r_2,r_3,r_4,s_2,s_3,s_4)$ (with midpoint $r_1$) induce an $S_{1,3,3}$ in $G$, which is a contradiction.    
Thus, $r_2$ and $s_2$ do not have any common neighbor in $N_1$, say $r_1r_2 \in E$, $s_1s_2 \in E$ with $r_1,s_1 \in N_1$, $r_1 \neq s_1$.

If $r_1$ and $s_1$ have a common neighbor $t_1 \in N_1$ then $t_0t_1 \in E$ with $t_0 \in N_0$. 
But then $(t_1,t_0,r_1,r_2,r_3,s_1,s_2,s_3)$ (with midpoint $t_1$) induce an $S_{1,3,3}$ in $G$, which is a contradiction.    
Thus, $r_1$ and $s_1$ do not have any common neighbor in $N_1$. 

\medskip

First assume that $r_1$ and $s_1$ have a common neighbor $r_0 \in N_0$. Recall that $r_1$ and $s_1$ do not have any common neighbor in $N_1$.

Next assume that $r_1$ or $s_1$ have some neighbor in $N_1$; without loss of generality, $s_1t_1 \in E$ with $t_1 \in N_1$ and $r_1t_1 \notin E$. 
Then $(s_1,t_1,r_0,r_1,r_2,s_2,s_3,s_4)$ (with midpoint $s_1$) induce an $S_{1,3,3}$ in $G$, which is a contradiction. Thus, $r_1$ as well as $s_1$ do not have any neighbor in $N_1$. 

Then $r_0$ must have distance at least $3$ with other $D$-neighbors in $N_0$, say $r_0t_1 \in E$ with $t_1 \in N_1$, $t_1 \neq r_1,s_1$. Thus, $t_1$ must have a  neighbor in $N_1$, say $t_1u_1 \in E$ with $u_1 \in N_1$. 
Recall that $r_2$ and $s_2$ do not have any common neighbor in $N_1$, i.e., $r_2t_1 \notin E$ or $s_2t_1 \notin E$; without loss of generality, $s_2t_1 \notin E$.

If $r_2t_1 \in E$ then $(t_1,u_1,r_2,r_3,r_4,r_0,s_1,s_2)$ (with midpoint $t_1$) induce an $S_{1,3,3}$ in $G$, which is a contradiction. 
Thus, $r_2t_1 \notin E$. But then $(r_0,t_1,r_1,r_2,r_3,s_1,s_2,s_3)$ (with midpoint $r_0$) induce an $S_{1,3,3}$ in $G$, 
which is a contradiction. 

Thus, $r_1$ and $s_1$ do not have any common neighbor in $N_0$, i.e., $r_0r_1 \in E$ and $s_0s_1 \in E$ with $r_0,s_0 \in N_0$, 
$r_0 \neq s_0$. Without loss of generality, assume that $dist_G(r_0,s_0)=4$, say $(r_0,r'_1,t_1,s'_1,s_0)$ induce a $P_5$ in $G[N_0 \cup N_1]$ 
(possibly $r_1=r'_1$ or $s_1=s'_1$).

If $r_1t_1 \in E$, $s_1t_1 \in E$ and $t_0t_1 \in E$ with $t_0 \in N_0$ then $(t_1,t_0,r_1,r_2,r_3,s_1,s_2,s_3)$ (with midpoint $t_1$) induce an $S_{1,3,3}$ 
in $G$, which is a contradiction. Thus, $r_1t_1 \notin E$ or $s_1t_1 \notin E$; without loss of generality, $r_1t_1 \notin E$. 

If $s_1t_1 \in E$ then $(t_1,t_0,r'_1,r_0,r_1,s_1,s_2,s_3)$ (with midpoint $t_1$) induce an $S_{1,3,3}$ in $G$, which is a contradiction.  
Thus, also $s_1t_1 \notin E$. But then $(t_1,t_0,r'_1,r_0,r_1,s'_1,s_0,s_1)$ (with midpoint $t_1$) induce an $S_{1,3,3}$ in $G$, which is a contradiction. 

\medskip

Then there is at most one $D$-vertex in $N_3 \cap X$. Analogously, there is at most one $D$-vertex in $N_3 \cap Y$, and Lemma \ref{atmostoneDN3XorY} is shown. 
\qed 

\medskip

Recall that by Lemma \ref{N5emptyN4indep}, $N_5=\emptyset$ and $N_4$ is independent, and by Lemma \ref{atmostoneDN3XorY}, 
there is at most one such $D$-vertex $r_3 \in D \cap N_3 \cap X$ and there is at most one such $D$-vertex $s_3 \in D \cap N_3 \cap Y$. Then one can update $D_{basis}:= D_{basis} \cup \{r_3,s_3\}$. 

\medskip

Then for every $t_3 \in N_3 \setminus \{r_3,s_3\}$ which does not contact $r_3$ or $s_3$, $t_3$ must have a $D$-neighbor $t_4 \in D \cap N_4$.  
If every neighbor of $t_3$ in $N_4$ contacts one of $r_3,s_3 \in N_3$, then it leads to a contradiction, and there is no such e.d.s.\ $D$ with $D_{basis}$. 
Now assume that $t_3t_4 \in E$ with $t_4 \in N_4$ and $r_3t_4 \notin E$, $s_3t_4 \notin E$. 
Then $t_3$ must have a $D$-neighbor $t_4 \in D \cap N_4$. Moreover, if $t_3t_4 \in E$ and $t_3t'_4 \in E$ with $t_4,t'_4 \in N_4$, $t_4 \neq t'_4$, 
$r_3t_4 \notin E$, $r_3t'_4 \notin E$, $s_3t_4 \notin E$, $s_3t'_4 \notin E$, then it leads to a contradiction, and there is no such e.d.s.\ $D$ with $D_{basis}$.
Thus, $t_3$ must have only one neighbor $t_4 \in N_4$ with $r_3t_4 \notin E$, $s_3t_4 \notin E$, and $t_4$ is $D_{basis}$-forced.

\medskip
 
Finally, the e.d.s.\ problem can be done in polynomial time and the proof of Theorem \ref{EDS133frbipgr} is done.
 
\medskip

\noindent
{\bf Acknowledgment.} The second author would like to witness that he just tries to pray a lot and is not able to do anything without that - ad laudem Domini.
 
\begin{footnotesize}

\end{footnotesize}

\end{document}